\documentclass[preprint2]{aastex63}

\newcommand{\myemail}{ricardo.rizzo@cab.inta-csic.es}
\newcommand{\kms}    {~km~s$^{-1}$}

\newcommand{\eg}     {e.\,g.\,}

\newcommand{\tastar} {$T_\mathrm{a}^*$}
\newcommand{\tmb}    {$T_\mathrm{MB}$}

\submitjournal{Astrophysical Journal Supplement Series}

\shorttitle{SiO, $^{29}$SiO, and $^{30}$SiO in oxygen-rich stars}
\shortauthors{Rizzo et al.}

\begin{document}


\title{SiO, $^{29}$SiO, and $^{30}$SiO emission from 67 oxygen-rich stars.\\
       A survey of 61 maser lines from 7 to 1\,mm}


\correspondingauthor{Ricardo Rizzo}
\email{\myemail}

\author{J. R. Rizzo}
\affil{Centro de Astrobiolog\'{\i}a (INTA-CSIC), 
       Ctra.~M-108, km.~4,
       E-28850 Torrej\'on de Ardoz, Madrid, Spain}
\affil{ISDEFE, Beatriz de Bobadilla 3, 
       E-28040 Madrid, Spain}

\author{J. Cernicharo}
\affil{Grupo de Astrof\'{\i}sica Molecular, 
       Instituto de F\'{\i}sica Fundamental (IFF-CSIC), 
       C/ Serrano 121, 
       E-28006 Madrid, Spain}

\author{C. Garc\'{\i}a-Mir\'o}
\affil{Joint Institute for VLBI ERIC (JIVE), 
       Oude Hoogeveensedijk 4, 
       7991 PD Dwingeloo, The Netherlands}


\begin{abstract}
Circumstellar environments of oxygen-rich stars are among the strongest SiO 
maser emitters. Physical processes such as collisions, infrared pumping and 
overlaps favors the inversion of level population and produce maser emission at 
different vibrational states. Despite numerous observational and theoretical 
efforts, we still do not have an unified picture including all the physical 
processes involved in the SiO maser emission. The aim of this work is to 
provide homogeneous data in a large sample of oxygen-rich stars. We present a 
survey of 67 oxygen-rich stars from 7 to 1 mm, in their rotational transitions 
from $J=1\rightarrow0$ to $J=5\rightarrow4$, for vibrational numbers $v$ from 0 
to 6 in the three main SiO isotopologues. We have used one of the 34\,m NASA 
antennas at Robledo and the IRAM 30\,m radio telescope. The first tentative 
detection of a $v=6$ line is reported, as well as the detection of new maser 
lines. The highest vibrational levels seem confined to small volumes, 
presumably close to the stars. The $J=1\rightarrow0$, $v=2$ line flux is 
greater than the corresponding $v=1$ in almost half of the sample, which may 
confirm a predicted dependence on the pulsation cycle. This database is 
potentially useful in models which should consider most of the physical agents, 
time dependency, and mass-loss rates. As by-product, we report detections of 27 
thermal rotational lines from other molecules, including isotopologues of SiS, 
H$_2$S, SO, SO$_2$, and NaCl.
\end{abstract}


\section{Introduction}

Evolved oxygen-rich stars are among the most powerful maser emitters known to 
date. Maser emission in $^{28}$SiO rotational transitions of vibrationally 
excited states is ubiquitous and very intense in these sources. Rotational 
transitions up to $J=7\rightarrow6$ have been detected in the vibrational 
states $v=1, 2, 3$ and 4 \citep[][and references therein]{jew87,gra95,cer93}. A 
certain number of maser lines have been also detected in the ground and excited 
vibrational states of the rare isotopologues $^{29}$SiO and $^{30}$SiO 
\citep[e.g.][]{deg83,cer91,alc92,cer92,gon96}.

Despite the large amount of observational data available in the literature, SiO 
maser emission has complex aspects that deserve further observational and 
theoretical effort. Besides the overall inversion of the rotational levels in 
each vibrational ladder, the SiO emission displays intriguing anomalies in some 
specific rotational lines, such as drastic changes in intensity from one 
rotational line to the next within the same $v$-state. Although the general 
inversion process seems to be relatively well understood 
\citep[e.g.][]{buj94a,buj94b}, the differences in the emission of adjacent 
rotational lines are difficult to explain upon standard radiative and 
collisional pumping models. 

Such anomalies, which are particularly important for the high-$v$ states of 
$^{28}$SiO, also apply to the less-abundant isotopologues $^{29}$SiO and 
$^{30}$SiO in $v=0, 1, 2$ and 3 \citep{cer91,cer92,cer93}. They have been 
interpreted as a result of infrared overlaps between the ro-vibrational lines 
of the $^{28}$SiO, $^{29}$SiO and $^{30}$SiO \citep{gon97,her00}. The 
excitation of the vibrational levels depends on the amount of dust in the 
envelope and on the effective temperature of the star photosphere. Furthermore, 
an overlap between SiO and H$_2$O has been proposed to explain the $v=2$ 
$J=2\rightarrow1$ line of $^{28}$SiO \citep{olo81,olo85,lan84,buj96}.

Optical pumping is also a possible mechanism to account for the observed 
features in these lines. \citet{rau96} have modelled synthetic atmospheres of 
M-stars immersed in expanding envelopes and show that optical pumping may 
account for some features of maser emission even without collisions; moreover, 
the regions of inversion may be different from one isotopologue to another, 
depending on the velocity difference between the photosphere and the emitting 
volume. Optical pumping through the A$^1\Pi$-X$^1\Sigma$ electronic transition 
could be very efficient in stars with moderate/low mass loss rate and hot 
photospheres in which photons at 250~nm could excite the $^1\Pi$ state. 
Excitation through the triplet states of SiO could also occur at 330~nm.

The excitation of the SiO masers through collisions with H$_2$ and He is a 
subject still open. Before the work of \citet{day06}, who performed new 
potential energy surface calculations of the system SiO/He/p-H$_2$, there were 
not accurate ro-vibrational collisional rates. Recently, \citet{bal17} extended 
the computation of collisional rates with He in the temperature range 300 -- 
6\,000~K, for the first six vibrational levels and up to the rotational states 
$J=40$.

Despite the many years since the discovery of SiO masers, robust and realistic 
models which take into account all the possible mechanisms for pumping are 
still necessary. To distinguish between the different physical processes 
leading to the population inversion of the SiO energy levels, and to retrieve 
key information about the physical conditions of the gas in the region between 
the photosphere and the dust growth zone, it is necessary to gather the most 
complete, simultaneous and homogeneous sample of the SiO rotational emission in 
the different vibrational states. The current availability of wideband backends 
spanning several GHz of instantaneous bandwidths make possible nowadays this 
kind of studies. These backends permit the simultaneous observation of multiple 
spectral lines avoiding the problem of the intrinsic variability. Furthermore, 
it permits to optimize observing time and to minimize the impact of the 
uncertainties of physical parameters associated to calibration and pointing.

There are dozens of SiO surveys of evolved stars, covering different lines, 
isotopologues and, in some cases, with great detail of the emitting regions 
thanks to the use of interferometers. Some of the most complete surveys include 
the observation of 11 lines of $^{28}$SiO in 6 evolved stars by \citet{sch82}, 
the line survey in the red supergiant \object{VY~CMa} by \citet{cer93}, the 
study of 12 sources by \citet{par98}, the survey by \citet{cho96} of six 
$J=1\rightarrow0$ transitions in various vibrational states of $^{28}$SiO and 
$^{29}$SiO, and the study of the $J = 1\rightarrow0$ $v=1$ and 2 lines done by 
\citet{kim10}. Most of the other studies, however, have been carried out in 
different epochs, with different instruments and, sometimes, without a complete 
coverage of the rotational transitions from different vibrational levels. Due 
to the variability of the SiO maser emission and to the complexity of the 
physics of stellar pulsation (which is in turn associated to the line pumping), 
the observations gathered in different epochs cannot be compared to infer 
robust conclusions about the excitation of SiO.

In this paper, we report a survey of SiO, $^{29}$SiO, and $^{30}$SiO maser 
lines towards 67 evolved oxygen-rich stars. The selected sources span different 
mass losses, temperatures, and C/O abundance ratio\,\footnote{The C/O abundance 
ratio is $<$1 for the whole sample. Carbon stars, which have C/O$>$1, exhibit 
SiO thermal emission but not maser lines.}. As a by-product, this work also 
provides valuable information about the molecular content of these objects in 
the surveyed frequency ranges. The survey was done using one of the 34~m 
antenna of the Madrid Deep Space Communications Complex (hereafter MDSCC) and 
the IRAM 30~m radio telescope at Pico Veleta. A total of 61 transitions from 
$J=1\rightarrow0$ to $5\rightarrow4$, and $v=0$ to 6 was observed in the whole 
sample. In Sect.~2 we describe the observations. We present the results 
in Sect.~3. In Sect.~4 we discuss some global results of the survey, including 
polarization, variability, some special cases and the identification of other 
spectral lines. We summarize the main findings in Sect.~5. In a follow up 
paper, we plan to use this database to model the circumstellar envelopes 
(CSEs), using an already developed non-local radiative transfer code which 
includes infrared overlaps between the three silicon isotopes of SiO, 
collisional pumping, and optical and infrared excitation.

\section{Observations}
\subsection{Overview, sources and strategy}
As said in the previous section, we have done a survey of 67 oxygen-rich 
evolved stars in their SiO, $^{29}$SiO, and $^{30}$SiO maser line emission. The 
\mbox{DSS-54} antenna of the Madrid Deep Space Communications Complex (MDSCC) 
was used in the first part of the survey to observe the $J=1\rightarrow0$ 
lines, at a wavelength of 7~mm ($\sim~43$~GHz). Based on these initial results, 
we observed the most relevant sources in the $J=2\rightarrow1$ to 
$5\rightarrow4$, at wavelengths of 3, 2 and 1~mm, using the 30~m IRAM radio 
telescope at Pico Veleta (Spain). The MDSCC observations were done between 
March and July 2012, and the IRAM observations in August 2012. 

We have chosen the observing modes and the bands in order to optimize the 
tuning of the largest possible number of simultaneous lines. The lines observed 
(61 in total) and their corresponding frequencies are indicated in Table 
\ref{freqs}. The frequencies have been obtained from the 
CDMS\footnote{{\tt http://www.astro.uni-koeln.de/cgi-bin/cdmssearch}} 
\citep{mul01,mul05} and the 
JPL\footnote{{\tt http://spec.jpl.nasa.gov/ftp/pub/catalog/catform.html}} 
\citep{pic98} catalogues. We also used information from MADEX
\footnote{{\tt https://nanocosmos.iff.csic.es/madex/}} 
\citep{cer12} when the above mentioned catalogues did not provided sufficient 
precision. The list of the observed sources are presented in Table 
\ref{sources}, together with some useful information, such as the date of 
observation, transitions, observing modes, velocity spacing, integration time, 
$rms$ noise, and polarization recorded. The \mbox{DSS-54} antenna is able to 
record both circular polarizations, while the 30~m radio telescope records 
lineal polarizations.

The half-power beam width (HPBW) of both telescopes at the different 
frequencies are indicated in Table \ref{effic}, together with sensitivities and 
efficiencies. The spectra have been corrected for atmospheric opacity and 
elevation gain (\tastar\ scale) during the observation, and converted to flux 
density ($S$ scale) during the reduction process. The conversion from \tastar\ 
to main-beam temperature (\tmb) can be done by 
$T_\mathrm{MB}=T_\mathrm{a}^*/\eta_\mathrm{MB}$, being $\eta_\mathrm{MB}$ the 
main beam efficiency, also depicted in Table \ref{effic}. Unless specifically 
stated, we use the flux density scale (in Jy) throughout all the paper, and in 
the online tables. 

\subsection{MDSCC observations}

We used the NASA \mbox{DSS-54} antenna of the \mbox{MDSCC} to observe the 
$J=1\rightarrow0$ lines included in the survey (Table \ref{freqs}). The \mbox
{DSS-54} is a beam-wave-guide antenna having 34\,m in diameter. The new Q-band 
receiver was employed, which has a temperature of 35~K at 43~GHz \citep{riz13}. 
The final system temperatures were between 90 and 130~K, in the antenna 
temperature scale.

The two circular polarizations from the receiver were processed by the new 
wideband backend \citep{riz12}. The observations were done in 14 different 
sessions between March 23 and July 26, 2012. At the beginning, the backend had 
the capacity to process a single circular polarization, with a bandwidth of 
1.5~GHz, 8192 channels, and a frequency spacing of $\sim183$~kHz (equivalent to 
1.28\kms). At some point during the survey, a second FFTS board were 
incorporated, allowing the possibility of observing both circular polarizations 
simultaneously and to use a high-resolution mode, which provided 500~MHz of 
bandwidth, 16384 channels, and a frequency spacing of $\sim31$~kHz (equivalent 
to 0.21\kms). The broad band FFTS backend was centered at 42.75~GHz, which 
allowed to simultaneously observe the 10 lines indicated in Table \ref{freqs}. 
The 500~MHz bandwidth FFTS was always centered at 42.97~GHz, in order to 
simultaneously observe the $J=1\rightarrow0$ maser lines of SiO $v=1, 2$ and 
$^{29}$SiO $v=0$. This setup is indicated as $1-0^*$ in Table \ref{sources}.

A total of 56 stars have been observed using this telescope with the largest 
bandwidth (1.5~GHz), 11 of those during two or more days for cross-checking and 
to study possible variability. A number of 28 out of those 56 target stars have 
also been observed with the high-resolution mode (i.e., 500~MHz bandwidth) to 
increase details about the line profiles.

We corrected the observed positions by assuming the standard pointing model for 
this antenna in Q-band, which is accurate to within 8\arcsec. This model was 
also regularly checked and improved at specific sessions in the epoch of 
observations, and we estimate that pointing errors can account intensity 
uncertainties always below 10\%. At the beginning of each observing session, 
we also cross-checked the pointing by means of the observation of sources 
included in the paper of \citet{riz12}. Focus was optimized for Q-band using 
specific calibration sessions, and automatically corrected during observations 
as part of the standard operational mode of the antenna.

Most of the observations were done in position-switching mode, with a reference 
located at 6\arcmin\ in azimuth. In 38 cases, we observed in 
frequency-switching mode, with a frequency throw of 13.2~MHz to avoid ripples. 
The atmospheric opacity was always between 0.07 and 0.1.

\subsection{IRAM observations}

We used the IRAM 30~m radio telescope in Pico Veleta, Spain, to observe the 
$J=2\rightarrow1$ to $5\rightarrow4$ lines (Table \ref{freqs}). The 
observations were done between 2012 August 1 and 4. Precipitable water vapor 
varied between 4 and 12~mm, which resulted in opacities at 225~GHz between 0.1 
and 0.7. Most of the observations were done during night time.

The EMIR \citep[Eight MIxer Receiver;][]{car12} was used for all the 
observations. The focal plane geometry of this receiver allows us to gather 2 
bands simultaneously, in two linear polarizations. We used the new FTS backend 
\citep{kle09} which provided a total of eight units having 4~GHz of bandwidth 
each and a frequency spacing of $\sim$~195~kHz. Four of these units were always 
used at the 3~mm band, tuned at 85.045~GHz (the central frequency of the 
$J=2\rightarrow1$ lines) and 88.775 GHz, in both polarizations. The other band 
(2 or 1~mm bands) was chosen according to weather conditions to ensure the best 
use of the observing time. Central frequencies were 128.675, 172.025, and 
215.575~GHz for the $J=3\rightarrow2$ to $5\rightarrow4$ bands. A complementary 
tuning at 168.290~GHz was necessary to include the $J=4\rightarrow3$ lines of 
SiO at $v=4$ to 6, $^{29}$SiO at $v=2$ to 5, and $^{30}$SiO at $v=0$ to 3; this 
additional mode is indicated as ``$4-3^*$'' in Table \ref{sources}. The 3~mm 
unit centered at 88.775~GHz was used to search of other molecules.

The observations were done in wobbler switching mode (which provides very flat 
baselines), with a throw at 90\arcsec\ at a rate of 2~Hz. System temperatures 
varied in the ranges 110--220~K, 134--360~K, 770--3080~K, and 390--1300~K for 
the $J=2\rightarrow1$ to $5\rightarrow4$ bands. 
The spectra have been automatically corrected by atmospheric opacity during the 
observations using the ATM package \citep{cer85, par01}. The antenna gain 
elevation curve, sensitivity and efficiencies provided in Table \ref{effic} 
were computed based on the observatory web pages. 

Pointing was checked by $1-1.5$~hr, and is accurate to within 3\arcsec. 
Every day, we also observed \object{W3(OH)} as a line calibrator \citep{mau89}. 
Focus was checked and corrected at the beginning of each session and after 
sunset. 

A total of 38 sources have been observed, 11 of them not included in the MDSCC 
sample.

\section{Results}
\subsection{Overview}

All detections are presented in Table \ref{detections}. In each row, the table 
depicts the observation ID, the source name, the transition detected, the 
parameters of the lines, and individual comments (if any). The transitions in 
each entry contains the isotopologue, the rotational quantum numbers 
$J\rightarrow J-1$, and the vibrational state. The parameters include the line 
flux ($F$), the flux-weighted velocity ($V_\mathrm{LSR}$), and the velocity 
range of emission above \mbox{3-sigma} ($V_\mathrm{min}$ and 
$V_\mathrm{max}$)\footnote{$V_\mathrm{min}$ and $V_\mathrm{max}$ were 
determined by the first and last occurrences of two consecutive channels with 
temperatures above 3-sigma.}. $F$ and $V_\mathrm{LSR}$ were computed in the 
velocity range ($V_\mathrm{min}, V_\mathrm{max}$). Some relevant information 
are also added in several entries to Table \ref{detections} as individual 
comments. These remarks are mostly referred to the cases when data smoothing 
has been applied to improve the signal-to-noise ratio. In other cases 
information about polarization and line blending is included. The only 
instrumental feature that we noted lies $\sim 20$~MHz apart from the 
$J=5\rightarrow 4$, $v=3$ line of $^{28}$SiO; although the detection and 
parameters are fairly well determined, the cases in which the spurious feature 
is observed are commented in the table.

For some cases, improvement of the signal-to-noise ratio was necessary to 
achieve clear detections; this was done by the average of data gathered during 
different days and/or from different polarizations. These cases are shown in 
Table \ref{averages}, in the same format as Table \ref{detections}.  These 
averaged spectra not only allowed the detection of lines, but also permitted us 
to identify wide components in at lest two cases: the $J=2\rightarrow1$ $v=0$ 
line of $^{29}$SiO in \object{S~Per} and in the $J=2\rightarrow1$ $v=0$ line of 
$^{28}$SiO in \object{IRC+60154} (see comments in Table \ref{averages}).

\begin{figure*}
\includegraphics[width=0.75\textwidth]{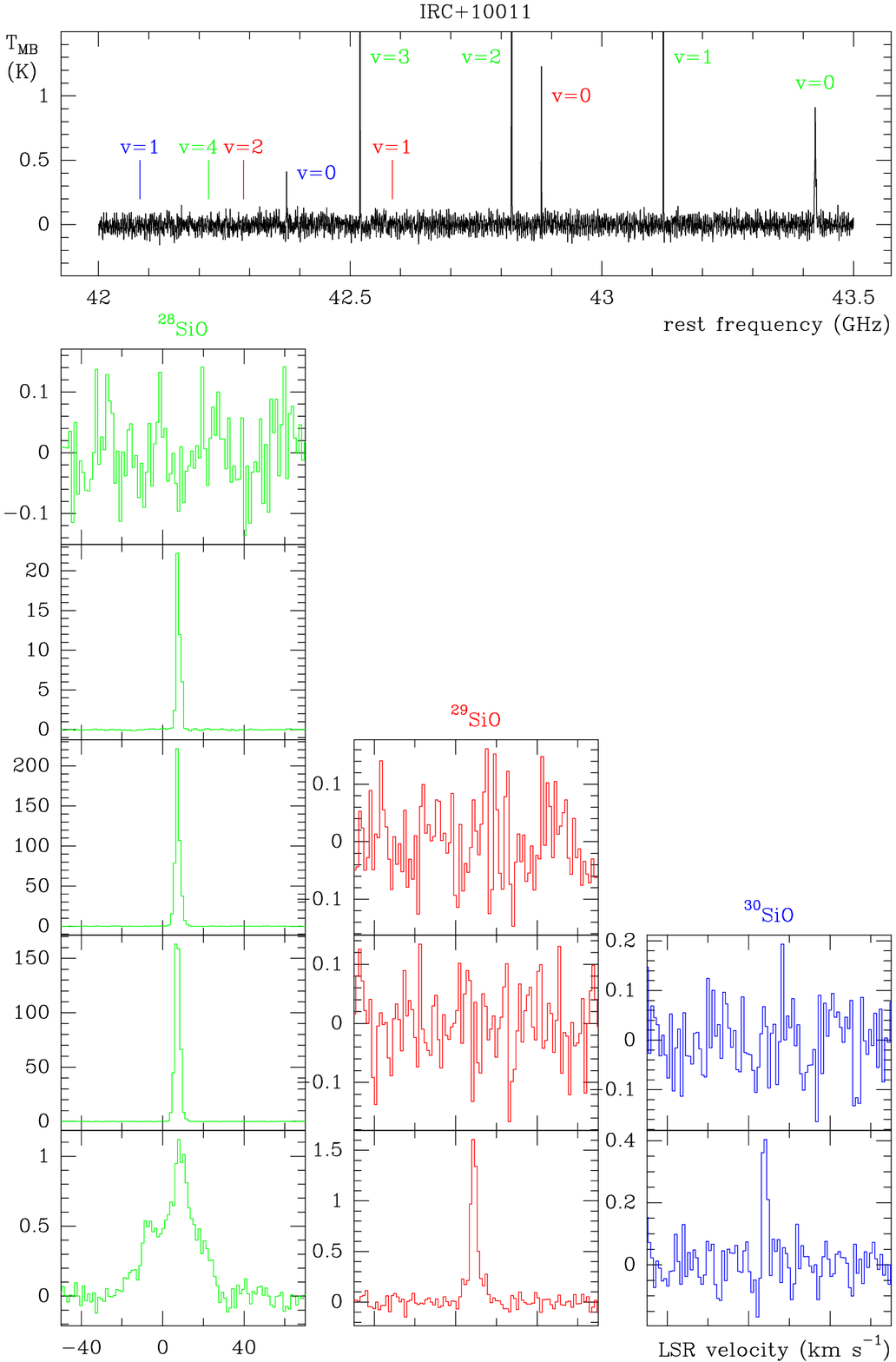}
\caption{A full range spectrum, aiming to show the complexity of lines. This 
example corresponds to the source IRC+10011, in the $J=1\rightarrow 0$ 
lines. The upper panel shows the whole spectrum, while the lower, small panels 
display individual lines. Green, red and blue correspond to $^{28}$SiO, 
$^{29}$SiO, and $^{30}$SiO, respectively. For each isotopomer, vibrational 
number $v$ increases from botton to top, starting at zero. In this example, six 
out of ten possible lines were detected.
\label{fullrange}}
\end{figure*}

As an example, the Fig.~\ref{fullrange} depicts the full range spectrum of one 
of the sources (\object{IRC+10011} = \object{WX~Psc}) in the $J=1\rightarrow0$ 
transitions. In this case, six over a total of ten possible lines are detected.

\subsection{Some examples of individual features}

In the following, a series of figures illustrates some of the properties found 
in the sample, particularly with respect to line shapes, variability and 
polarization.

\noindent{\bf Line profiles}. The physical conditions to pump masers are so 
restrictive that the emitting volumes are relatively small; therefore, one of 
the fingerprints of the maser emission is that the velocity components are 
narrow, typically 1\kms\ or even less. 

Depending on the pumping mechanisms and the physical conditions, the maser 
emitting regions dramatically change from one source to another, and also from 
one line to another \citep{gra09}. This trend is confirmed 
by high angular resolution observations \citep[see, for 
example,][]{sor07,wit07}. When these kind of sources are observed by single 
dishes, though, the different emitting regions and physical conditions are 
reflected in a variety of line profiles.

\begin{figure*}
\includegraphics[width=0.95\textwidth]{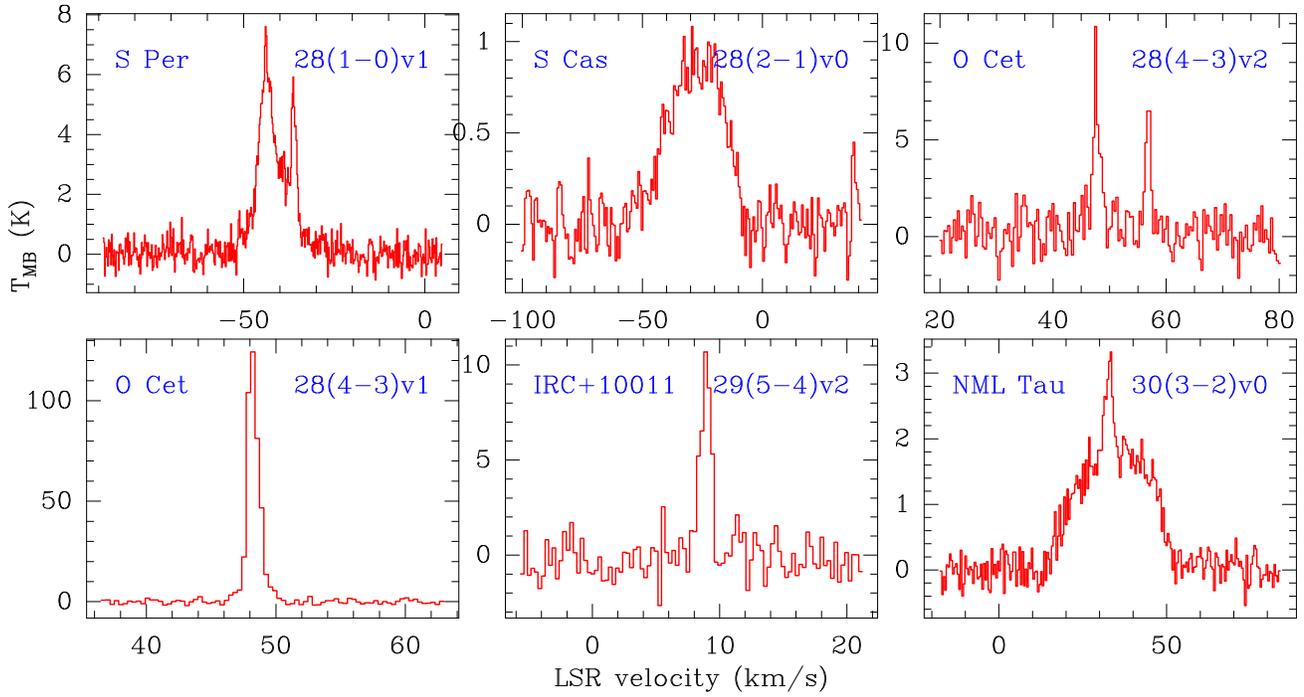}
\caption{A sample of six different line profiles, representative of the 
catalogue. Source name are indicated at the upper left corner of each panel, 
while the lines (abridged) are shown in the upper right corner.
\label{profiles}}
\end{figure*}

Figure \ref{profiles} shows six representative examples of the line shapes 
present in the catalog. For each panel, source name is indicated on the upper 
left and the transition (in abridged format) in the upper right corner. In the 
case of \object{S~Per} the spectrum is dominated by two peaks, but is the 
result of a superposition of several individual blended components. In the case 
of \object{S~Cas}, the shape of a truncated parabolic and wide line is 
indicative of thermal emission from the CSE as a whole. In 
\object{O~Cet} the $^{28}$SiO $J=4\rightarrow 3$, $v=2$ line displays a 
component at $\sim57$\kms\ which is right outside the velocity range of the 
thermal line; the $J=4\rightarrow 3$, $v=1$ line (lower left panel) has a 
Gaussian shape, and is centered at the star velocity. The line depicted for 
IRC+10011 is representative of a typical very narrow maser line. And finally, 
the line displayed for \object{NML~Tau} ($^{30}$SiO $J=3\rightarrow 2$, $v=0$) 
seems the result of the superposition of both the CSE thermal component and at 
least one maser component.

\begin{figure}
\includegraphics[width=0.95\textwidth]{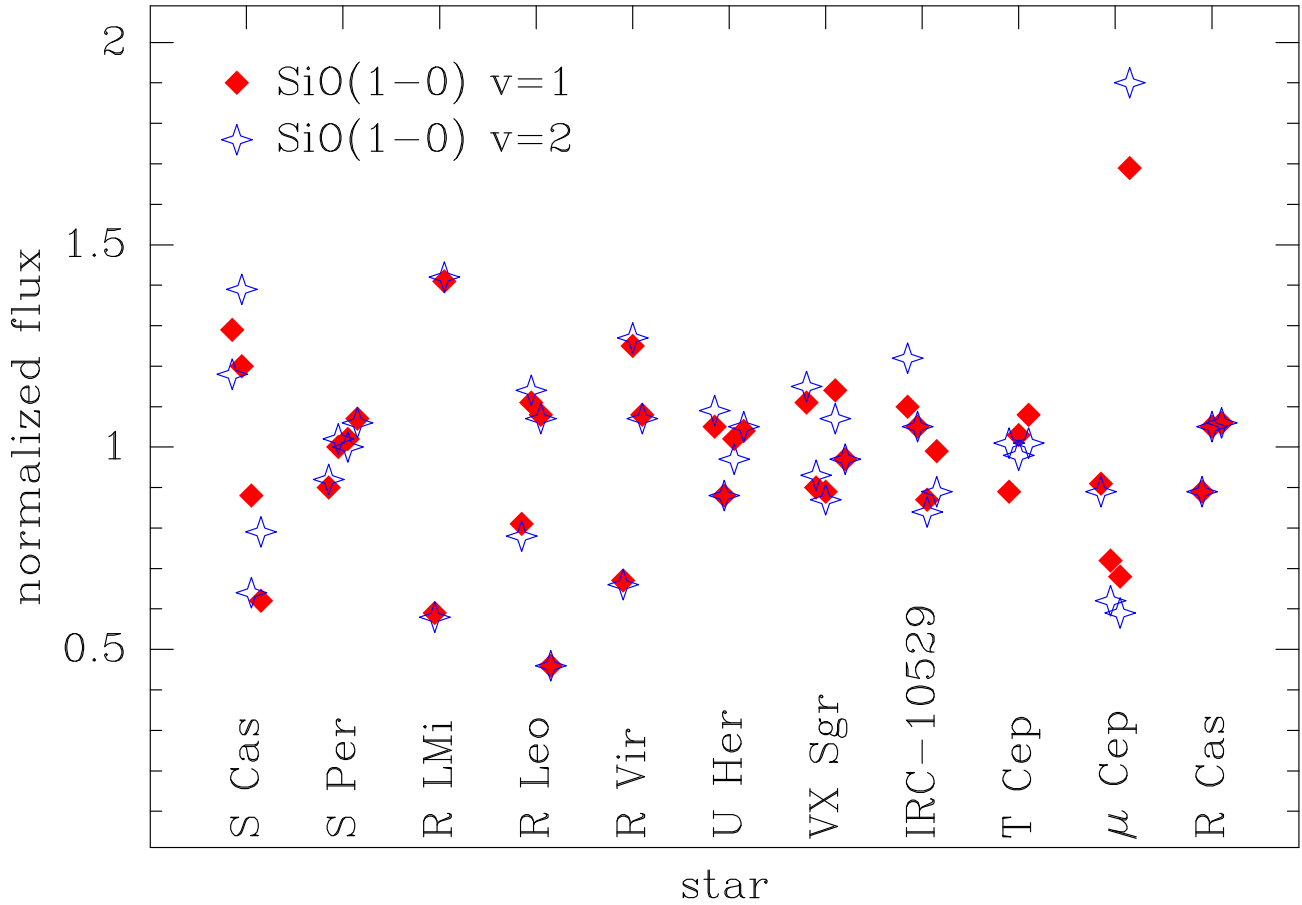}
\caption{
Variability of the $J=1\rightarrow 0$, $v=1$ and $v=2$ line fluxes 
corresponding to the eleven stars observed in two or more different days. 
Fluxes are normalized to their respective averages. Red and blue marks 
represent fluxes corresponding to the $v=1$ and $v=2$ lines, respectively. In 
most cases the variability is clearly noted on time scales of several days to 
few weeks. The variability of $\mu$~Cep is remarkable, where the highest fluxes 
in both lines are almost two times the other values (see text and 
Fig.~\ref{mucep}). 
\label{variab10}}
\end{figure}

\begin{figure}
\includegraphics[width=0.95\textwidth]{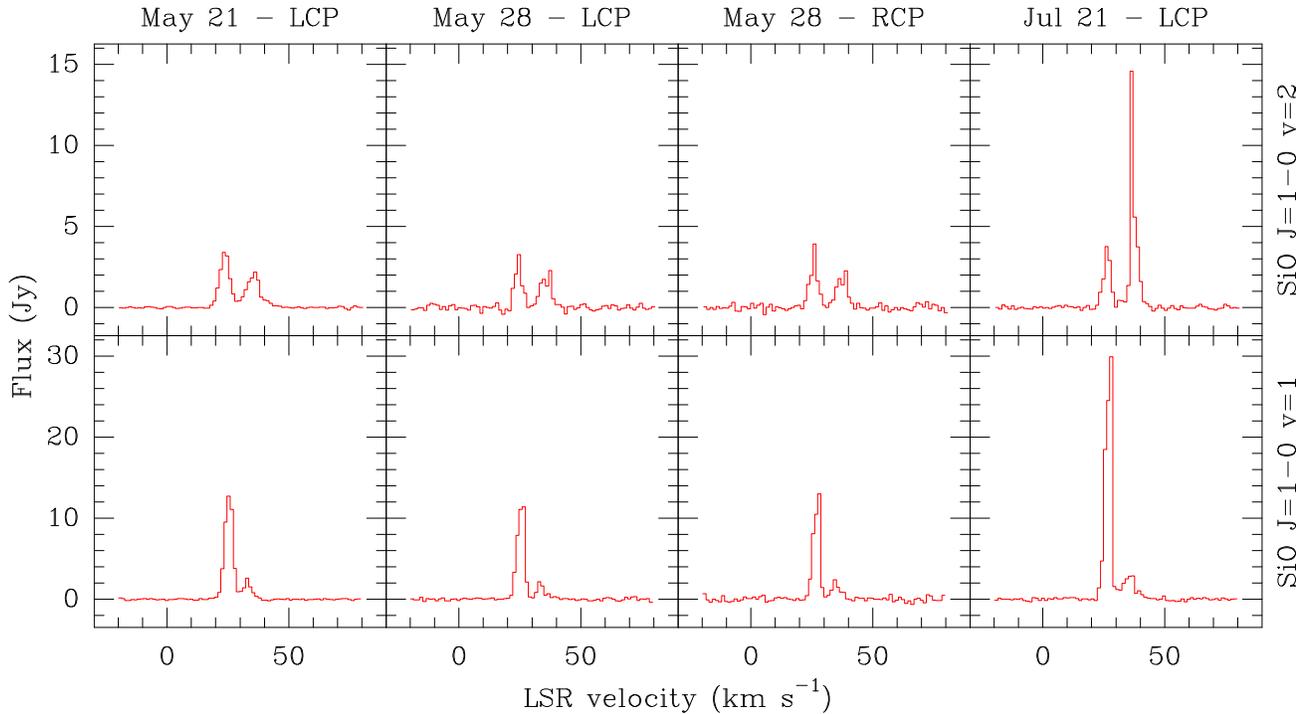}
\caption{
Variability of $\mu$~Cep. Lower and upper rows correspond to the 
$J=1\rightarrow 0$, $v=1$ and $v=2$ lines, respectively. Observation dates and 
circular polarizations are indicated on top. The emission does not seem 
significantly polarized (second and third columns). The notable increment 
experienced in the last day (fourth column) arise from different velocity 
components in the two maser lines.
\label{mucep}}
\end{figure}

\noindent{\bf Variability}. Most of the SiO and isotopologues maser lines in 
evolved stars develop a high degree of variability on scales from days to years 
\citep{hum02,par04}. 
The $J=1\rightarrow 0$ line observations performed with the Robledo antenna 
have been repeated in 11 sources, taking from 2 to 5 spectra spread in time 
from several days to few weeks, as indicated in Table~\ref{sources}. 
The Fig.~\ref{variab10} plots all the measured line fluxes, normalized to their 
respective averages. While some sources do not display large changes in any of 
the two lines, other sources vary significantly within the observed time frame. 
The standard deviation of these data is 22 and 26\% for the $v=1$ and $v=2$ 
lines, respectively. It is remarkable the case of $\mu$~Cep, which presents a 
dramatic variation where the highest fluxes roughly double the lowest ones. In 
order to provide more details, Fig.~\ref{mucep} shows all the observed spectra, 
gathered during three different days and in both circular polarizations. The 
emission is not significantly polarized, as we may infer from the second and 
third columns. It is evident in the figure that the line fluxes experienced a 
sudden increase in the last day, just some weeks after a rather stable maser 
emission; more interestingly, the changes are found in totally different 
velocity components for the $v=1$ and $v=2$ lines. 

Another clear example of line variability 
is shown in Fig.~\ref{variability}, where the $^{30}$SiO $J=1\rightarrow 0$, 
$v=0$ line of \object{VX~Sgr} significantly changed in just two weeks.

\noindent{\bf Polarization}. A significant part of the maser lines are often 
linearly and circularly polarized due to intrinsic magnetic fields 
\citep[e.g.][]{shi04,vle11,shi17}. The IRAM spectra contains valuable 
information about lineal polarization of these sources, although it is not 
possible to derive the Stokes parameters with the present observations. Even 
though, the simultaneous observations of both lineal polarizations during four 
consecutive nights allowed the discovery of highly polarized components. Two 
examples are shown in Fig.~\ref{polariz}: in \object{IRC+10011}, the $^{30}$SiO 
$J=2\rightarrow 1$ $v=0$ line depicts in the horizontal polarization a 
component at $\sim10$\kms) which is virtually absent in the vertical 
polarization; the second example is \object{VY CMa}, where the SiO 
$J=4\rightarrow 3$ $v=3$ line displays significantly differences between the 
two linear polarization in the two principal velocity components.

\subsection{Identification of other spectral lines}

For a significant fraction of the sample, a frequency range of up to 
$\approx$\,30~GHz has been surveyed. This large bandwidth permits also the 
search for other molecules. For each source, we averaged the spectra 
corresponding to all dates and polarizations, and looked for molecular species 
other than SiO and its isotopologues. 

A total of 27 lines have been detected. Table \ref{others_identify} provides 
the list of those spectral lines together with some useful information, such as 
the frequencies, quantum numbers, and energies of the upper levels. For the 
sake of brevity, we labeled the spectral lines by a letter followed by a 
number; the letter indicates the molecular species, while the number designates 
the transition detected, ordered by isotopologue and frequency. 

Detections are presented in the Table \ref{others_detections}. A total of 20 
stars with detections of some of the above mentioned 27 thermal lines are 
included. In all cases spectra have been smoothed to a velocity resolution of 
2\kms. Positive detections are labeled by ``Y'', negative results by ``N'', and 
detections after further smoothing by ``S''.

A brief analysis of the results are presented in Sect.~4 below.

\section{Discussion}
\subsection{The highest vibrationally excited lines}

\begin{figure*}
\includegraphics[width=0.95\textwidth]{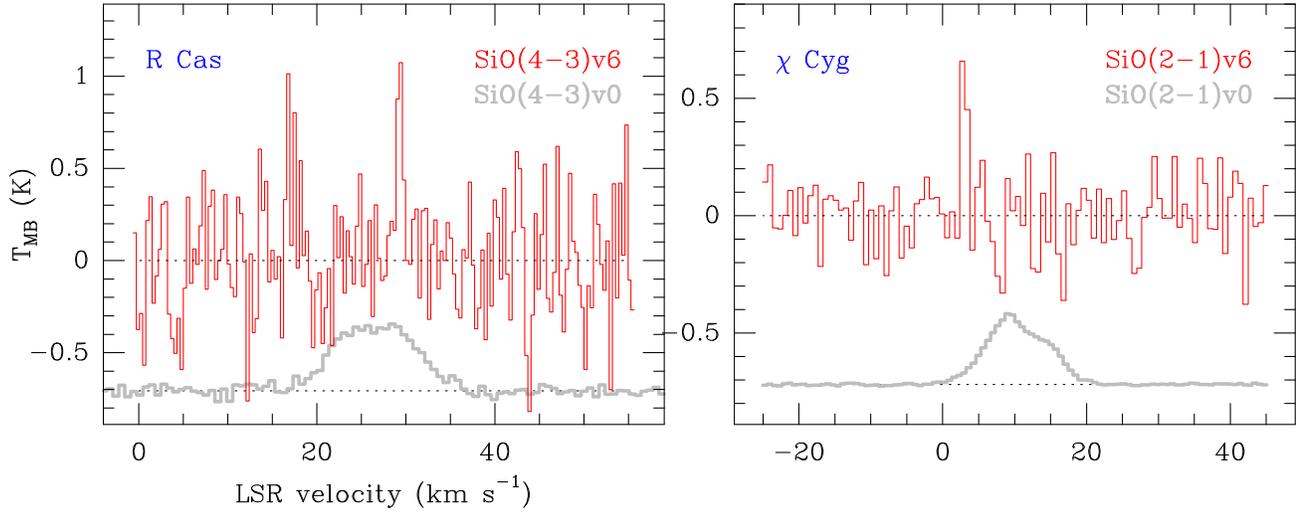}
\caption{First tentative detections of $v=6$ SiO maser lines. They correspond 
to \object{R~Cas} ($J=4\rightarrow 3$; left panel) and \object{$\chi$~Cyg} 
($J=2\rightarrow 1$; right panel). Both $v=6$ lines are very narrow and 
displayed in red. To visualize the velocity range of emission of the other SiO 
lines, the corresponding $v=0$ line (not to scale) are displayed in grey.
\label{v6}}
\end{figure*}

\object{R Cas} is one of the strongest SiO maser emitter, and has been the 
subject of numerous single-dish and interferometric observations up to $v=3$ 
\citep[see, for example][]{phi01, phi03, mci08, mci10, ass11, ass13}. We report 
here the first tentative detection of a $v=6$ line, corresponding to the 
rotational transition $J=4\rightarrow3$. The line, displayed in Fig.~\ref{v6} 
(left panel), is very narrow and has been significantly detected after 
averaging the two lineal polarizations (Table~\ref{averages}).

We also report the tentative detection of another $v=6$ line, the one 
corresponding to $J=2\rightarrow1$ line in \object{$\chi$ Cyg} (ID 348 in Table 
\ref{detections}). The line, very narrow, is also depicted in Fig.~\ref{v6} 
(right panel). \object{$\chi$ Cyg} also displays the only $v=5$ line of the 
whole survey. This also corresponds to the $J=2\rightarrow1$ line (ID 345 in 
Table \ref{detections}). The $J=2\rightarrow1$ lines have been observed on 
almost consecutive days (August 1, 2, and 4; scans from 345 to 350), and both 
highly vibrationally excited lines have been detected in only one scan. Taken 
into account that both lines are detected at $\approx 3\,\sigma$ level, 
\object{$\chi$ Cyg} deserves further observations, because both highly 
vibrationally excited lines may be polarized and rapidly variable, as has been 
suggested by recent observations of the intense $J=2\rightarrow1$, $v=1$ line 
\citep{gom20}. 

\begin{figure}
\includegraphics[width=0.95\textwidth]{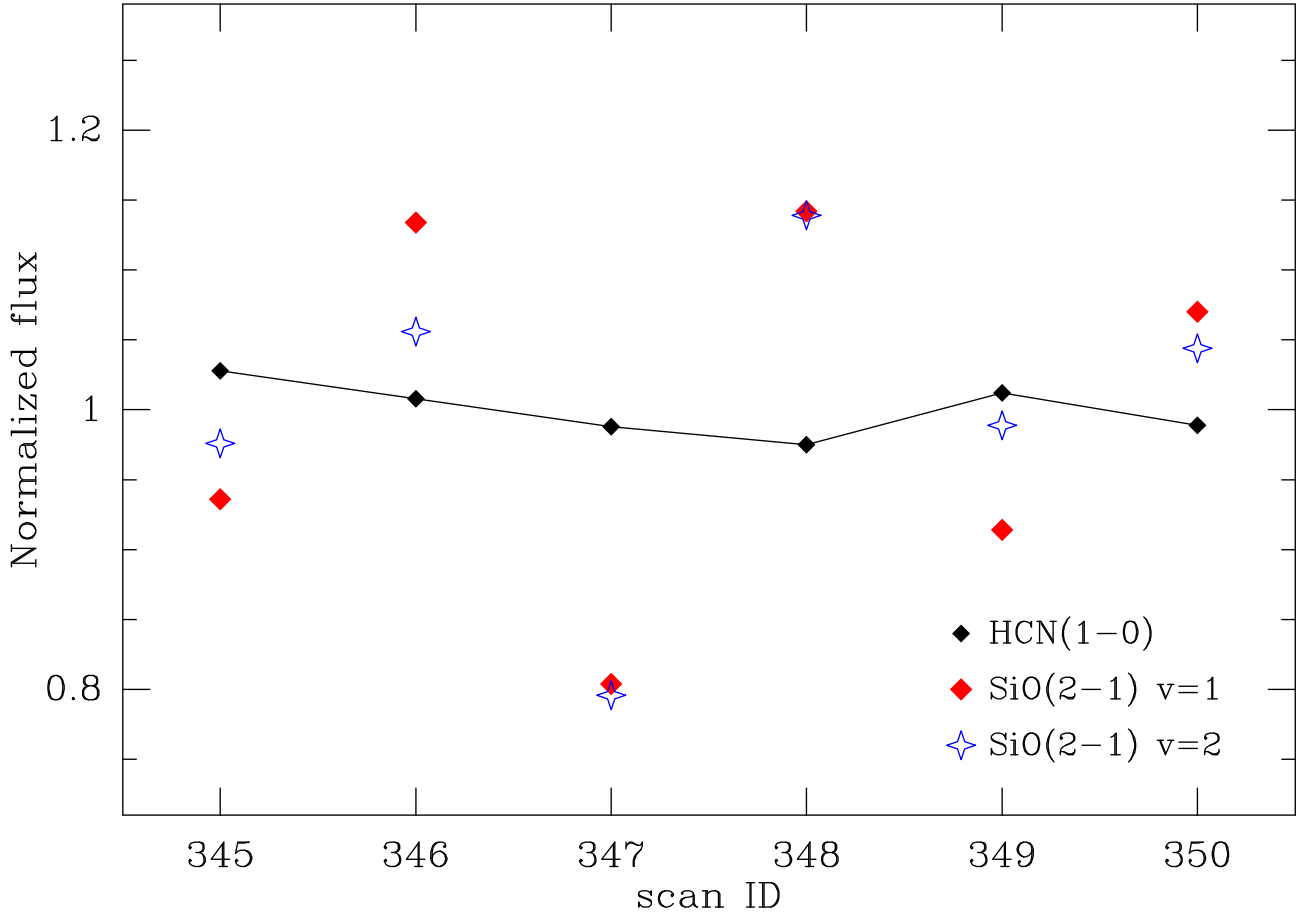}
\caption{Intraday variability of \object{$\chi$~Cyg}. Fluxes of the thermal HCN 
$J=1\rightarrow 0$ and the $J=2\rightarrow 1$, $v=1,2$ of SiO lines are plotted 
as a function of their scans IDs. To facilitate the comparison, fluxes 
(computed from $-5$ to $+25$~\kms) are normalized to their respective average 
values.  Scan IDs are explained in Table \ref{sources}, and correspond to both 
linear polarizations observed during August 1, 2, and 4. Note that the HCN 
normalized flux (black marks) remains well within $\pm3\%$, while the SiO line 
fluxes (blue and red marks) change significantly (up to $20\%$). This 
demonstrates the high polarization and rapid variability of the SiO lines in 
this source.
\label{intraday}}
\end{figure}

In \object{$\chi$ Cyg}, the intraday variability claimed by \citet{gom20} is 
confirmed by our data in the $v=1$ and $v=2$, $J=2\rightarrow1$ lines. The HCN 
$J=1\rightarrow0$ line, included in the same six scans as the $J=2\rightarrow1$ 
SiO lines, does not vary by more than $3\%$, while the $v=1$ and $v=2$ line 
fluxes are very dispersed around the mean value, with departures from the 
average up to $20\,\%$. This is clearly illustrated in Fig.~\ref{intraday}, 
where the integrated fluxes of the three lines are plotted as a function of the 
scan IDs; the fluxes are integrated between $-5$ and $+25$~\kms and normalized 
to the average of the six scans (345 to 350).

To our knowledge, there is only one reported detection of a $v=5$ line, which 
is that corresponding to $J=8\rightarrow7$ in \object{VY CMa} \citep{kam13}, 
using the SMA interferometer.

The $J=11\rightarrow10$, $v=4$ line has been reported in the high-mass young 
stellar object Orion Source~I \citep{hir18, kim19}. In an evolved star, 
however, there is only one detection of a $v=4$ line SiO maser: the 
$J=5\rightarrow4$ line in \object{VY CMa} \citep{cer93}. We detected the 
vibrationally excited $J=3\rightarrow2$ $v=4$ line in \object{VY CMa} and for 
the first time in other four sources: \object{IRC+10011}, \object{R Leo}, 
\object{VX Sgr}, and \object{S Per}. As expected, the line is very narrow and 
highly polarized, being \object{VY CMa} the only source where the line was 
detected in both lineal polarizations; the significant detection in 
\object{S Per} was reached after averaging both polarizations. The narrowness 
of the lines is not unusual, but the significance of the detections should be 
monitored carefully. It is worth noting that the maser emission of this line 
was predicted by \citet{her00} as the result of infrared line overlaps.

\subsection{About the emitting region}

Masers request different physical conditions to invert the level population. It 
is therefore expected that the emitting volumes change from one maser to 
another. This is confirmed by interferometry at low $v$-states 
\citep[\eg][]{gon10,kam13,ric13}. Such restrictive physical conditions are met 
through different mechanisms (radiative pumping, collisional pumping, 
overlaps), as explained in Sect.~1. In addition, we are dealing with pulsating 
stars, which adds the time dependency of such conditions.

\begin{figure}
\includegraphics[width=0.95\textwidth]{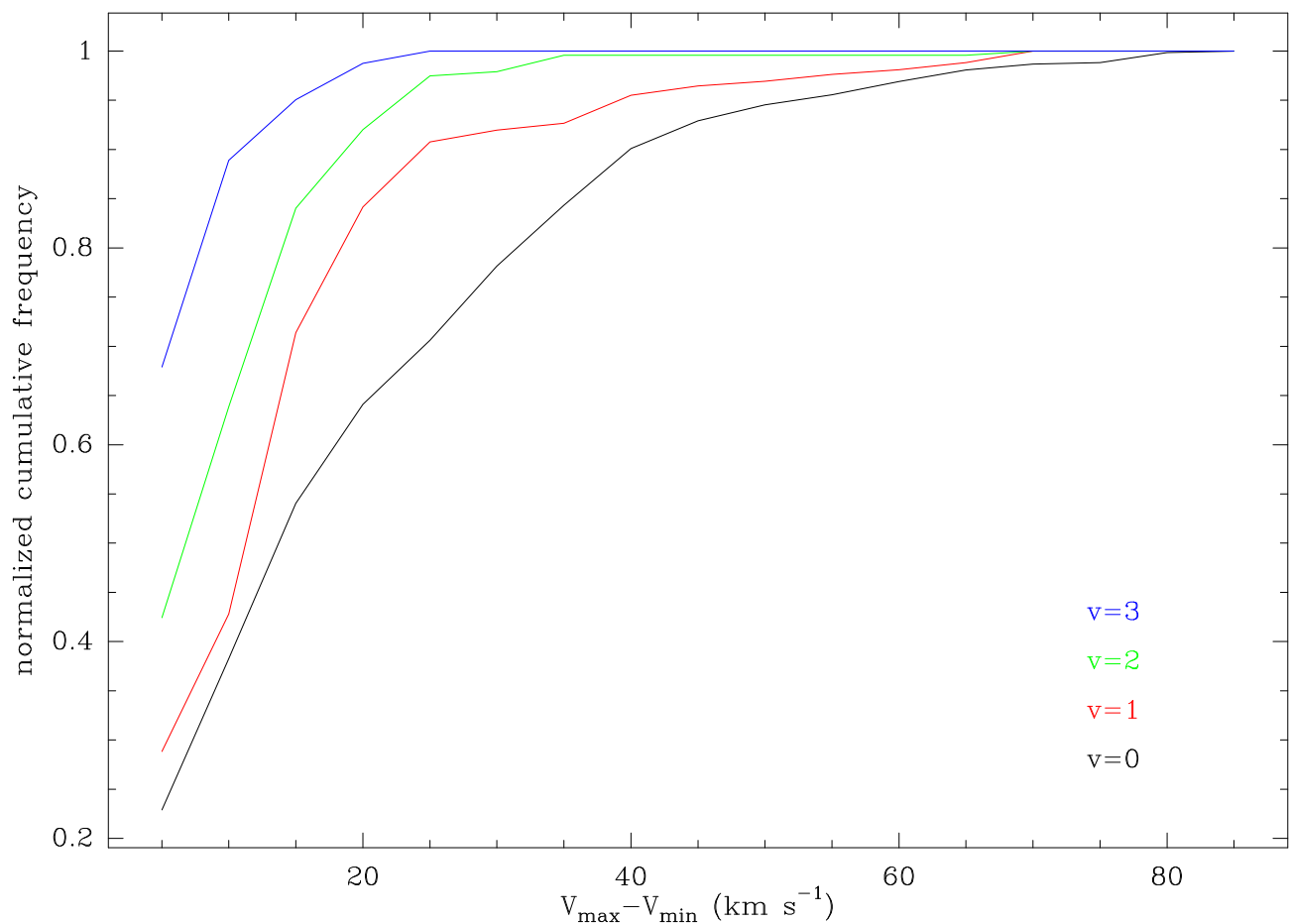}
\caption{Normalized cumulative frequencies of the different vibrational levels 
for the whole sample, as a function of the velocity range of emission.
\label{cum}}
\end{figure}

As our survey contains almost simultaneous observations of all masers, it is 
particularly suitable to get an idea about the overall distribution of the 
emitting regions. A first approach is provided in Fig.~\ref{cum}, where we plot 
the normalized cumulative frequencies of the vibrational levels 0 to 3, as a 
function of the velocity range of emission. As $v$ increases, the lines are 
more concentrated to smaller velocity ranges. Assuming that the velocity 
dispersion is correlated with the emitting volume, this tendency 
strongly suggests 
that the emitting region is confined to smaller volumes for larger 
$v$. As expected, this result also indicates that the physical conditions 
(temperature and density, IR radiative field) to produce maser emission become 
more restrictive as $v$ increases.

\citet{yun12} simulated SiO maser emission in Mira-type stars under non-LTE 
conditions, considering different velocity gradients, and covering a complete 
pulsation cycle. One of the most robust results is the prediction of the 
$J=1\rightarrow 0, v=2$ line being more intense than the corresponding $v=1$ in 
half of the pulsation cycle, and the opposite in the other half. We can test 
this prediction with our sample, because a reasonably high number of stars have 
been observed with co-occurrence of both lines. 

\begin{figure}
\includegraphics[width=0.95\textwidth]{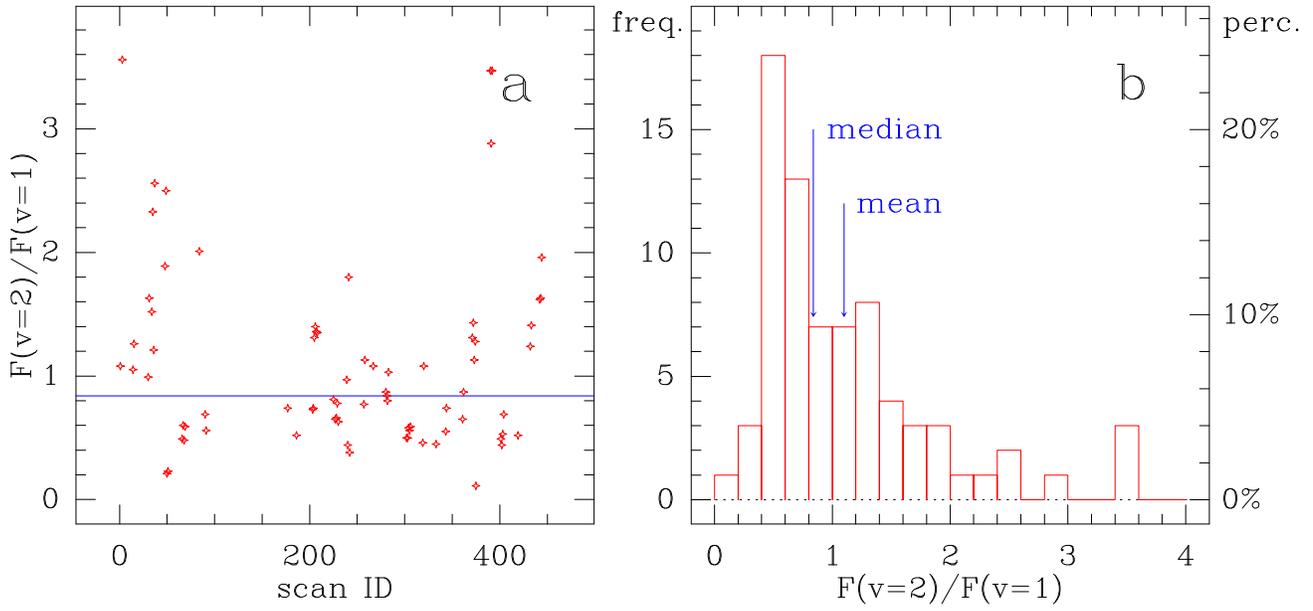}
\caption{
Relative intensities of the $J=1\rightarrow 0$, $v=1$ and $v=2$ lines. (a) Flux 
ratios $v=2/v=1$ plotted as a function of the scan ID; blue line indicates the 
median of the sample. (b) Distribution of the same ratios, expressed in 
absolute (left axis) and relative (right axis) frequencies; the median (0.84) 
and the mean (0.11) of the sample are also plotted as blue arrows.
\label{freqs_v1v2}}
\end{figure}

Our data contain a total of 150 detections of both lines (75 each) 
corresponding to 38 stars, and other four cases: two with detections of the 
$v=1$ line and other two with detections of the $v=2$ line. The 
Fig.~\ref{freqs_v1v2} sketches the main findings. In the left panel 
--Fig.~\ref{freqs_v1v2}(a)--, we plot the flux ratio of the $v=2$ to $v=1$ 
lines as a function of the scan ID, while in the right panel 
--Fig.~\ref{freqs_v1v2}(b)--, it is shown the distribution of the same line 
ratio in bins of 0.2 width. The median (0.84) and mean (0.11) of the sample are 
also indicated in the figure, which helps to infer a rather uniform 
distribution of the ratios around values close to one, i.e., without a clear 
dominance of one line over the other.

We have counted a total of 33 sources (44\%) where F($v=2$) is greater than 
F($v=1$), and 42 cases (56\%) with the opposite behavior. Overall results of 
this rather simple analysis seem to confirm the prediction made by 
\citet{yun12}, although a more thorough and case-by-case analysis should be 
performed to provide a firm confirmation.

\subsection{``Bonus'' thermal lines}

In order to provide an overall view about the thermal lines detected, we 
divided them into four groups. The first group is constituted by the C-bearing 
molecules HCO$^+$, HNC, HCN and H$^{13}$CN. HCO$^+$ (A1 in Table 
\ref{others_identify}) is a wide spread molecule besides H$_2$ and CO, abundant 
in a variety of astronomical sources such as comets, diffuse clouds, and 
molecular clouds, but with abundances below the predicted values in AGB stars 
\citep{gla96}. HCO$^+$ was firstly discovered in \object{VY~CMa} \citep{ziu07} 
and later in other sources \citep[e.g.][]{pul11}, including TX~Cam and NML Cyg. 
We detect HCO$^+$ in 10 sources.

HNC (B1 in Table \ref{others_identify}) was first tentatively reported by 
\citet{lin88} in TX Cam, and later detected by \citet{ziu09} in 
\object{VY~CMa}. We detect HNC in \object{TX~Cam}, \object{NML~Tau}, and 
\object{NML~Cyg}, but not in \object{VY~CMa} probably due to a high noise 
level.
HCN (C1 in Table \ref{others_identify}) is the most common molecule found in 
the survey, detected in 19 out of the 20 sources. It was already detected by 
several authors \citep[e.g.][]{lin88,ner89,ziu09} in some of the stars of our 
sample (IRC+10011, NML~Tau, TX~Cam, VY~CMa, VX~Sgr, NML~Cyg, and R~Cas). Its 
isotopologue H$^{13}$CN (C2 and C3 in Table \ref{others_identify}) is less 
abundant and therefore hardly detected; it was reported and analyzed in 
IRC+10011, NML~Tau (= \object{IK~Tau}), VY~CMa and NML~Cyg 
\citep{ner89,ten10a,vel17}. 

The second group (lines D and E in Table~\ref{others_identify}) is constituted 
by the sulfur-bearing molecules SiS and H$_2$S, and some of their isotopologues. 
Around evolved stars, SiS and H$_2$S are tracers of warm gas, probably above 
100\,K \citep[][and references therein]{san15}, and are good tracers of the 
dust formation zones \citep{cer11}. SiS was firstly reported by \citet{lin88} 
in TX~Cam, and later in NML~Tau \citep{buj94c}; we confirm here those 
detections and add other six to the list: \object{IRC+10011}, \object{S~Cas}, 
\object{VY~CMa}, \object{VX~Sgr}, \object{V111~Oph}, and \object{NML~Cyg}.

H$_2$S was firmly detected by \citet{omo93} in IRC+10011, NML~Tau, VY~CMa, and 
NML~Cyg; a tentative detection was also reported by \citet{dan17} in 
\object{V1111~Oph}. We add here the detection of H$_2$S in \object{TX~Cam}.

The third group (lines F and G in Table~\ref{others_identify}) is formed by 
sulfur oxides. Together with the SiO maser lines, sulfur oxides are the most 
abundant molecules in oxygen-rich CSEs. SO (F1 to F5 lines in Table 
\ref{others_identify}) and SO$_2$ (G1 to G5 in the same table) have been 
detected since the first molecular studies of these sources. \citet{omo93} 
detected up to three lines of SO$_2$ in IRC+10011, NML~Tau, VY~CMa, RX~Boo, and 
NML Cyg, none of them coincident to those reported in our survey. Later, 
\citet{ten10a} performed a sensitive survey of VY CMa at 1mm and reported 
several S-bearing molecules in this source. A similar spectral survey towards 
\object{NML Tau}, but with a high spectral coverage (from 79 to 356\, GHz) was 
recently published \citep{vel17} with similar findings. We detected SO and 
SO$_2$ in the same sources as in the \citet{omo93} article, and also in VX~Sgr 
and R~Cas. The less abundant isotopologue $^{34}$SO (line F5) is detected only 
in \object{NML Cyg}. The $16_{3,13}-16_{2,14}$ line of SO$_2$ (G5), with the 
highest upper energy level (147.8~K) is only detected in \object{VY Cma} and 
\object{NML Cyg}.

The fourth group is constituted by NaCl and its isotopologue Na$^{37}$Cl. Four 
lines have been clearly identified, labeled as H1 to H4 in Table 
\ref{others_identify}. Since the first identification in CSEs \citep{cer87}, 
this metal refractory molecule has been firmly observed in \object{IK~Tau} and 
\object{VY CMa} \citep{mil07,ten10b,kam13,dec16,vel17}; recently, it was 
tentatively detected in \object{R~Dor} \citep{deb18}, although not confirmed 
with ALMA data \citep{dec18}. We failed to detect NaCl in \object{IK~Tau} and 
\object{VY CMa}, although report the first detection in \object{NML Cyg}, as 
shown in Fig.~\ref{nacl}. Besides the lines quoted in Table 
\ref{others_detections}, the observed frequencies include also the 
$J=10\rightarrow 9$ and $13\rightarrow12$ lines of Na$^{37}$Cl (also shown in 
Fig.~\ref{nacl}), not detected probably due to insufficient noise level.

\section{Conclusions}

This work reports 
the results of a nearly complete survey of SiO, $^{29}$SiO, and $^{30}$SiO 
emission for $J=1\rightarrow 0$ to $5\rightarrow 4$, in 67 oxygen-rich stars. 
The stars have been chosen to span a large range of mass-loss rates, from 
10$^{-8}$ up to 10$^{-4}$ M$\odot$ yr$^{-1}$. In all rotational transitions, we 
surveyed simultaneously the vibrational levels $v$ from 0 to 6, completing a 
list of 61 maser lines. A total of 1474 detections is reported. 

The observations were made in a relatively short time (weeks for the 
$J=1\rightarrow 0$ and four days for the others); therefore, we can consider 
that most stages of the pulsation phases were randomly tested.

\begin{figure}
\includegraphics[width=0.95\textwidth]{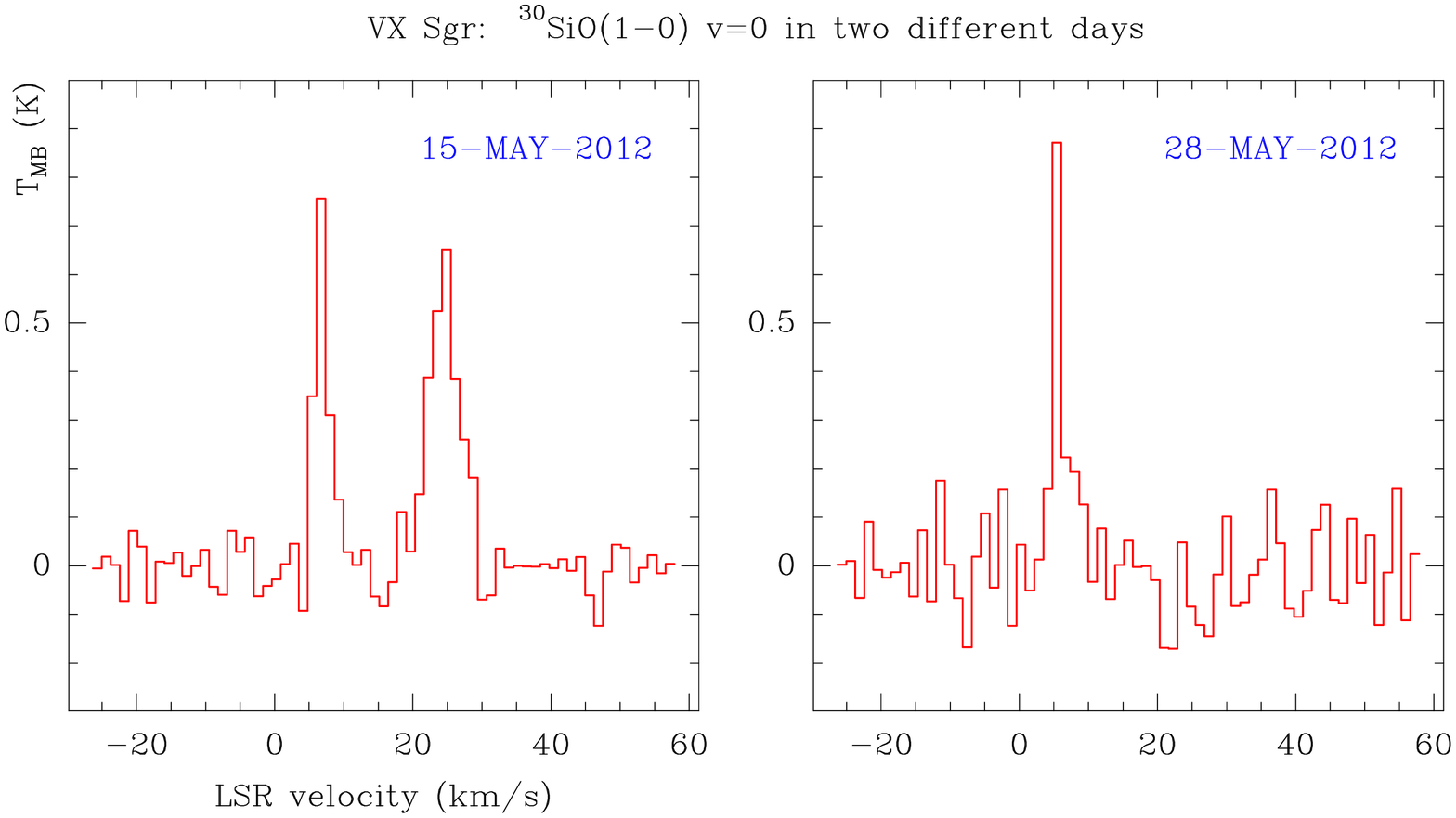}
\caption{An example of variability. The $^{30}$SiO $J=1\rightarrow 0$, $v=0$ 
line of \object{VX~Sgr} has been observed two times, with a time separation of 
only two weeks, but resulting in very different spectra.
\label{variability}}
\end{figure}

\begin{figure*}
\includegraphics[width=0.95\textwidth]{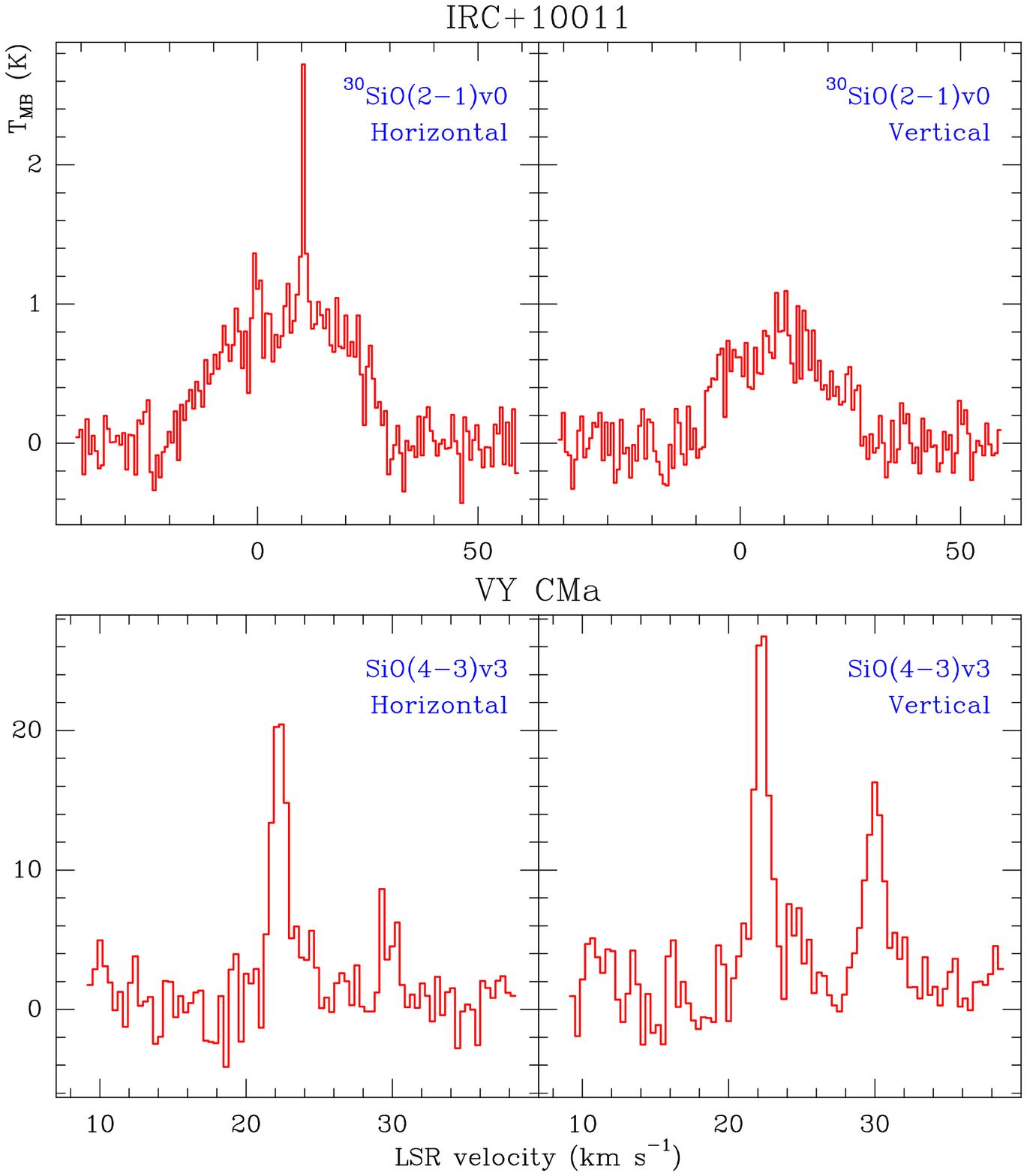}
\caption{Two examples of highly linearly polarized velocity components. The 
cases shown are the $^{30}$SiO $J=2\rightarrow 1$, $v=0$ line towards 
\object{IRC+10011} and the $^{28}$SiO $J=4\rightarrow 3$, $v=3$ line towards 
\object{VY CMa}.
\label{polariz}}
\end{figure*}

As expected, several maser lines exhibit significant variability and are 
highly linearly or circularly polarized. The most prominent cases have been 
highlighted (Sect.~3.2, Figs.~\ref{variability} and \ref{polariz}). 

Several lines are reported for the first time. It is remarkable the first 
detection of a $v=6$ line in \object{R Cas} and \object{$\chi$ Cyg}. $\chi$~Cyg 
also displays the only $v=5$ line detected. The $v=4$ vibrational state only 
depicts rotational lines in the $J=3\rightarrow2$ transition; very narrow and 
highly polarized, this line was detected in VY~CMa and for the first time in 
IRC+10011, R~Leo, VX~Sgr, and S~Per.

\begin{figure*}
\includegraphics[width=0.8\textwidth]{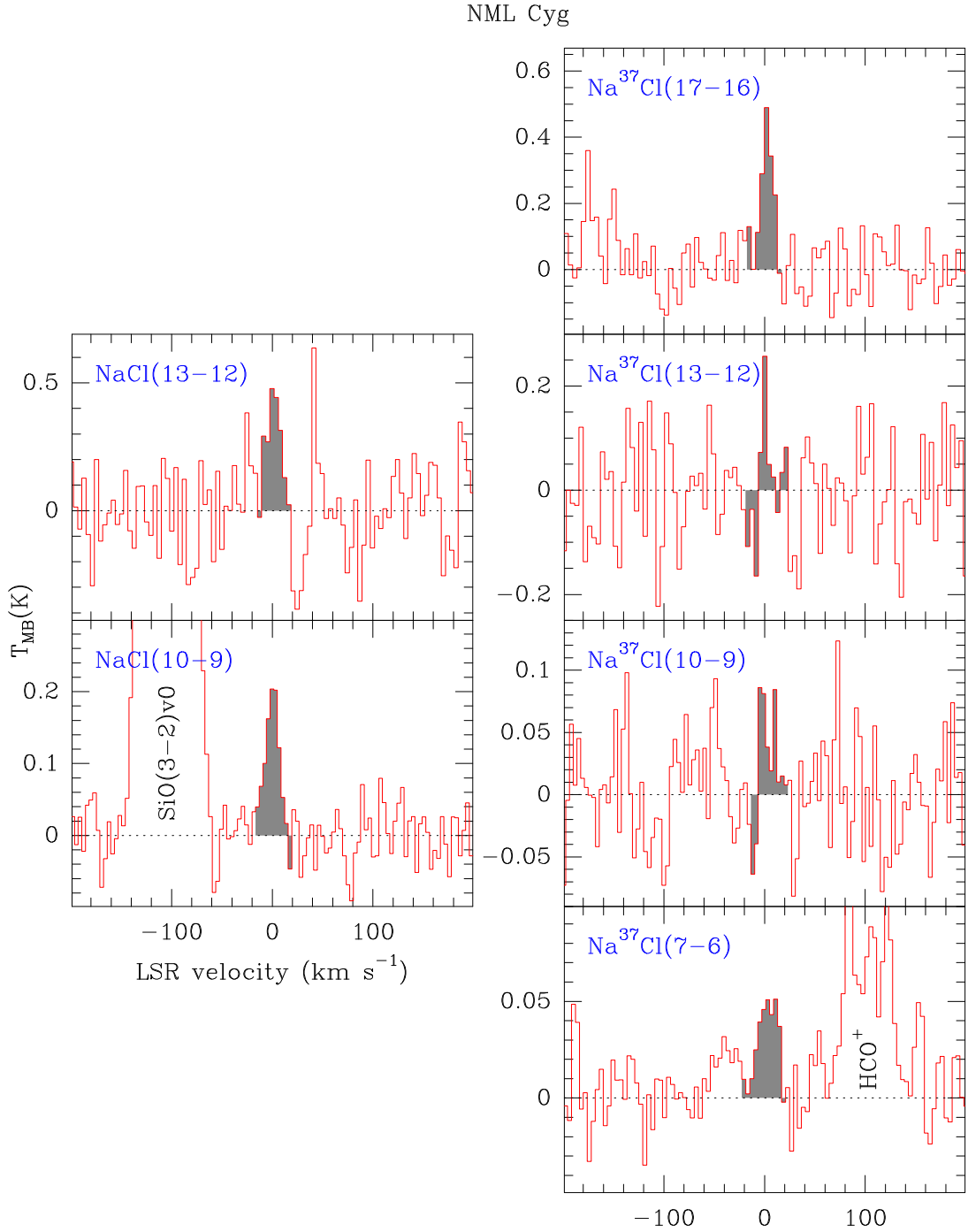}
\caption{NaCl and Na$^{37}$Cl lines in \object{NML Cyg}. Transitions are 
indicated at the top left of each panel. Lines are shaded in the velocity range 
of emission of other lines in this source. Two other close lines (HCO$^+$ and 
SiO) fall inside the plot, as indicated. 
\label{nacl}}
\end{figure*}

As a by-product, we also report the detection of other 27 thermal lines over a 
total of 20 sources. The lines correspond to common density tracers (like 
HCO$^+$ and HCN), refractory molecules (like SiS), S-bearing molecules (H$_2$S, 
SO, SO$_2$) and the less observed NaCl, detected for the first time in NML~Cyg 
(Fig.~\ref{nacl}).

SiO plays a key role in the process of dust formation under the appropriate 
physical conditions. The database generated by this survey would be the basis 
of ambitious modeling of SiO maser emission and the overall evolution of the 
circumstellar envelopes.



\acknowledgments
This work was partially done under the Host Country Radio Astronomy program at 
MDSCC. The authors wish to thank the MDSCC and IRAM staffs for their kind and 
professional support during the observations. 
J.R.R. acknowledges the support from projects ESP2017-86582-C4-1-R and PID2019-105552RB-C41 (Ministerio de Ciencia e Innovaci\'on). 
J.C. thanks ERC for Synergy grant ERC-2013-Syg-610256-NANOCOSMOS.



\facilities{MDSCC:DSS-54, IRAM:30m}


\clearpage

\clearpage
\begin{table*}
\centering
\caption{Frequencies of the observed lines (in MHz)}
\label{freqs}

\begin{tabular}{crrrrr}
\multicolumn{6}{c}{SiO}\\
\hline
$v$ & $J=1\rightarrow0$ & $J=2\rightarrow1$ & $J=3\rightarrow2$ & $J=4\rightarrow3$ & $J=5\rightarrow4$ \\
\hline
 0 & 43423.85163 & 86846.9600 & 130268.6100 & 173688.3100 & 217104.9800 \\
 1 & 43122.07310 & 86243.3700 & 129363.2400 & 172481.1500 & 215595.9500 \\
 2 & 42820.58624 & 85640.4600 & 128458.8872 & 171275.2800 & 214088.5400 \\
 3 & 42519.38333 & 85038.0464 & 127555.2739 & 170070.3500 & 212582.6000 \\
 4 & 42218.45653 & 84436.1911 & 126652.4901 & 168866.6332 & 211077.8700 \\
 5 &             & 83834.8731 &             & 167663.9945 &             \\
 6 &             & 83234.0760 &             & 166462.3977 &             \\
\\
\multicolumn{6}{c}{$^{29}$SiO}\\
\hline
$v$ & $J=1\rightarrow0$ & $J=2\rightarrow1$ & $J=3\rightarrow2$ & $J=4\rightarrow3$ & $J=5\rightarrow4$ \\
\hline
 0 & 42879.94655 & 85759.1990 & 128637.0500 & 171512.8027 & 214385.7577 \\
 1 & 42583.82840 & 85166.9603 & 127748.6912 & 170328.3228 & 212905.1554 \\
 2 & 42287.99473 & 84575.2900 & 126861.1851 & 169144.9799 & 211425.9744 \\
 3 &             & 83984.1744 &             & 167962.7457 & 209948.1793 \\
 4 &             &            &             & 166781.6000 &  \\
 5 &             &            &             & 165602.4900 &  \\
\\
\multicolumn{6}{c}{$^{30}$SiO}\\
\hline
$v$ & $J=1\rightarrow0$ & $J=2\rightarrow1$ & $J=3\rightarrow2$ & $J=4\rightarrow3$ & $J=5\rightarrow4$ \\
\hline
 0 & 42373.42369 & 84746.1702 & 127117.5479 & 169486.8766 & 211853.4736 \\
 1 & 42082.54453 & 84164.4095 &             & 168323.3529 & 210399.0665 \\
 2 &             & 83583.2041 &             & 167160.9396 &             \\
 3 &             &            &             & 165999.6090 &             \\
\hline
\end{tabular}
\vspace{5mm}
\end{table*}
\vspace{5mm}

\clearpage

\begin{longrotatetable}
\begin{deluxetable}{rlccrcccrrrc}
\centering
\tablecaption{Sources observed\label{sources}}
\tablewidth{0pt}
\tablehead{
\colhead{ID} & \colhead{Source} & \colhead{RA (J2000)} & \colhead{Dec (J2000)} & \colhead{$V_\mathrm{LSR}$} & 
\colhead{Date\tablenotemark{a}} & \colhead{$J\rightarrow\,J-1$} & \colhead{Mode\tablenotemark{b}} & \colhead{vel.~spacing} & 
\colhead{$t_{\mathrm int}$} & \colhead{$rms$\tablenotemark{c}} & \colhead{Pol.\tablenotemark{d}}\\
\colhead{} & \colhead{} & \colhead{hh:mm:ss.ss} & \colhead{$\pm$dd:mm:ss.s} & \colhead{\kms} & 
\colhead{mm-dd} & \colhead{} & \colhead{} & \colhead{\kms} & 
\colhead{min.} & \colhead{Jy}}

\startdata
1	 & 	Y Cas	 & 	00:03:21.40	 & 	$+$55:40:52.2	 & 	$-$17.0	 & 	05-21	 & 	       $1-0$	 & 	  FSw	 & 	1.28	 & 	24	 & 	0.05	 & 	LCP	\\
2	 & 	IRC+40004	 & 	00:06:53.24	 & 	$+$43:05:03.0	 & 	$-$19.7	 & 	05-21	 & 	       $1-0$	 & 	  FSw	 & 	1.28	 & 	9	 & 	0.06	 & 	LCP	\\
3	 & 	T Cas	 & 	00:23:14.27	 & 	$+$55:47:33.2	 & 	1.0	 & 	05-21	 & 	       $1-0$	 & 	  FSw	 & 	1.28	 & 	15	 & 	0.06	 & 	LCP	\\
4	 & 		 & 						 & 				 & 		 & 	08-01	 & 	       $2-1$	 & 	  WSw	 & 	0.66	 & 	16	 & 	0.14	 & 	HLP	\\
5	 & 		 & 						 & 				 & 		 & 	08-01	 & 	       $2-1$	 & 	  WSw	 & 	0.66	 & 	16	 & 	0.14	 & 	VLP	\\
6	 & 		 & 						 & 				 & 		 & 	08-03	 & 	       $2-1$	 & 	  WSw	 & 	0.66	 & 	30	 & 	0.12	 & 	HLP	\\
7	 & 		 & 						 & 				 & 		 & 	08-03	 & 	       $2-1$	 & 	  WSw	 & 	0.66	 & 	30	 & 	0.12	 & 	VLP	\\
8	 & 		 & 						 & 				 & 		 & 	08-03	 & 	       $3-2$	 & 	  WSw	 & 	0.46	 & 	15	 & 	0.14	 & 	HLP	\\
\enddata

\tablecomments{Table \ref{sources} is entirely published in the electronic 
edition of the {\it Astrophysical Journal Supplement Series}.  A portion is 
shown here for guidance regarding its form and content.}
\tablenotetext{a}{\ All dates in year 2012.}
\tablenotetext{b}{\ PSw: position switching. FSw: frequency switching. WSw 
                  Wobbler switching.}
\tablenotetext{c}{\ 1-$\sigma$ rms value.}
\tablenotetext{d}{\ Polarizations: LCP~=~Left circular; RCP~=~Right circular; 
                  HLP~=~Horizontal linear; VLP~=~Vertical linear.}
\tablenotetext{e}{\ $1-0^*$ includes only the $v=1$ and 2 lines of $^{28}$SiO, 
                  and the $v=0$ line of $^{29}$SiO.}
\tablenotetext{f}{\ $4-3^*$ includes the $v=4$ to 6 lines of $^{28}$SiO, 
                  the $v=2$ to 5 lines of $^{29}$SiO,
                  and the $v=0$ to 3 lines of $^{30}$SiO.}

\end{deluxetable}
\end{longrotatetable}

\clearpage
\begin{table}
\begin{center}
\caption{Efficiencies and intensity conversions}
\label{effic}

\end{longrotatetable}

\end{document}